\documentclass[twocolumn,twoside,slac_two,]{revtex4}
\usepackage{graphicx}
\usepackage{fancyhdr}
\newcommand{\lsim}{\lower.7ex\hbox{$
    \;\stackrel{\textstyle<}{\sim}\;$}}

\begin{document}

\title{Precision Searches for Physics Beyond the Standard Model}

%

\author{B. Lee Roberts}
\affiliation{Department of Physics, Boston University, Boston, MA 02215, USA}

\begin{abstract}
The ``precision'' frontier, which is closely related to the ``intensity'' 
frontier, provides a complementary path to the discovery of physics beyond
the Standard Model.  Several examples of discoveries that would change our
view of the physical world are: Charged lepton flavor violation, e.g. 
muon electron conversion; the discovery of a permanent electric dipole moment
of the electron, neutron, muon, or a nucleus.  In this paper I focus mostly on
phenomena meditated by a dipole interaction, including the anomalous magnetic
moment of the muon.  
\end{abstract}

\maketitle

\thispagestyle{fancy}


\section{Introduction}

Precision measurements have played an important role in our understanding of
the subatomic world.  The discovery of anomalous magnetic moments is one early
example, where the hyperfine structure of hydrogen (HHFS) was found to be too
large for the standard theory to be correct~\cite{blr-HHFS}.  
The {Dirac equation}~\cite{blr-Dirac28}
\begin{equation}
i\left(\partial_{\mu}-ieA_{\mu}\left(x\right)\right)\gamma^{\mu}\psi\left(x\right)
=m_{e}\psi\left(x\right),
\label{eq:dirac}
\end{equation}
predicted a magnetic dipole moment (MDM) for the electron 
\begin{equation}
\vec{\mu}_{e}=g_{e}\left( \frac{Qe}{2m_{e}}\right) \vec{s}
\label{eq:1.2}
\end{equation}
with the factor $g_e \equiv 2$ (and $e>0$). The increase in the 
hydrogen hyperfine levels could be interpreted as coming from an
additional magnetic moment. 
Motivated by the HHFS dilemma,  Schwinger~\cite{blr-schwinger}
 carried out the first ``loop''
calculation, and predicted that the electron had an additional
(anomalous) magnetic moment
\begin{equation}
a_e = \frac{\alpha}{2\pi} \ {\rm where}\ a_e = \frac {(g_e-2)}{2} .
\end{equation} 
  The subsequent
precision spectroscopy measurements of Kusch and
 Foley~\cite{blr-Kusch} obtained a
measurement of $g_e$ that was in good agreement with Schwinger's prediction.  

In 1950, Purcell and Ramsey suggested that an electric dipole moment (EDM) would
violate parity invariance {\sl P}, and proposed to search for 
the neutron electric dipole
moment~\cite{blr-Purcell}. This was of course the correct New-Physics effect  to
 look for, but in the wrong
place. Their initial experiment~\cite{blr-smith57} achieved a limit of 
$\vert d_n \vert < 5 \times 10^{-20}\, e\cdot$cm, a 
null result which has been pushed 
down to $2.9 \times 10^{-26}\, e\cdot$cm during the subsequent fifty-some years.
It was realized in 1957~\cite{blr-landau57,blr-ramsey58}
 that an EDM would also violate time-reversal
symmetry, {\sl T}, and by implication {\sl CP}.  Presumably, new, as yet
undiscovered sources of {\sl CP} violation are responsible for the
matter-antimatter asymmetry in the universe, and would partially explain why
we are here.

Other important examples of precision measurements
 are the search for charged lepton flavor
violation (neutrino mixing having already been discovered), precision M{\o}ller
scattering, neutron beta decay, rare or forbidden kaon decays, and the
precision measurements of the muon anomalous magnetic moment.

In this paper I focus on magnetic and electric dipole moments, and on searches
for charged lepton flavor violation (CLFV) in the muon sector, which also may
go though a dipole interaction.  Related papers at this conference
 are those by Tom Browder on 
{\sl CP} violation and by Dave Hitlin, who
discussed searches for CLFV in the $\tau$ sector. 

\section{The Dipole Operators}

As mentioned above, the Dirac equation is inadequate to describe the measured
magnetic moment of the electron.  It is necessary to add a ``Pauli'' term
\begin{equation}
 \frac{Qe}{4m_{e}}a_{e}F_{\mu\nu}(x)\sigma^{\mu\nu}\psi(x)
\label{eq:blr-pauli}
\end{equation} 
which in modern language is a dimension 5 operator that must arise from
loops in a renormalizable theory.  New Physics (NP) can also contribute
through loops, with $a({\rm NP})= C(m/\Lambda)^2$ where 
$C \simeq {\mathcal O}(1)$, or $\simeq {\mathcal O}(\alpha)$ in
weak coupling loop scenarios.
In the same spirit, one could add the following Pauli-like term
\begin{equation}
\frac{i}{ 2}
d_{e}F_{\mu\nu}\left(x\right)\sigma^{\mu\nu}\gamma_{5}\psi\left(x\right)
\label{eq:EDM}
\end{equation}
which represents the electric dipole moment interaction, where
\begin{equation}
\vec d = \eta \left( \frac{Qe}{2mc}\right) \vec s.
\end{equation}
and the quantity $\eta$ plays the role for the EDM that $g$ plays for the
MDM. One way to parameterize the effects of NP on $a$ and $d$ is by
$d(NP) = a(NP)(e/2m)\tan \phi^{NP}$~\cite{blr-CzarLM}.

The electromagnetic current is given by
\begin{equation}
\left\langle f\left(p'\right)\left|J_{\mu}^{em}\right|f\left(p\right)\right\rangle =\bar{u}_{f}\left(p'\right)\Gamma_{\mu}u_{f}\left(p\right)\label{eq:2.1}\end{equation}
where $\bar{u}_{f}$ and $u_{f}$ are Dirac spinor fields and $\Gamma_{\mu}$ has
the general Lorentz structure
\begin{eqnarray}
\Gamma_{\mu} &=& 
F_{1}\left(q^{2}\right)\gamma_{\mu}
+iF_{2}\left(q^{2}\right)\sigma_{\mu\nu}q^{\nu}
-F_{3}\left(q^{2}\right)\sigma_{\mu\nu}q^{\nu}\gamma_{5}
\nonumber \\ 
&&
+F_{A}\left(q^{2}\right)\left(\gamma_{\mu}q^{2}-2m_{f}q_{\mu}\right)\gamma_{5}; 
\label{eq:2.2}
\end{eqnarray}
with $F_1(0) = Qe$ the electric charge, $F_2(0) = a(Qe/2m)$ the anomalous 
magnetic moment, and $F_3 = dQ$ the electric dipole moment.  I will ignore
the last term, the anapole moment.

The anomalous part of the dipole moment interaction
\begin{equation}
\bar u_{\mu}\left[Qe F_1(q^2)\gamma_{\beta} +
{i Qe \over 2m_{\mu}}F_2(q^2)\sigma_{\beta \delta}q^{\delta}\right]u_{\mu}
\label{eq:blr-dipole}
\end{equation}
connects states of opposite helicity, i.e. it is chiral changing, giving
 it a unique sensitivity to NP interactions, e.g. the sensitivity to 
$\tan \beta$ in supersymmetric (SUSY) theories. In most SUSY models,
the contribution to $a_\mu$ depends on the SUSY mass scale, the sign of the 
$\mu$ parameter, and $\tan \beta$. A simple SUSY model
with equal masses~\cite{blr-CzarLM,blr-Czar01} gives the SUSY contribution as:
\begin{equation}
\simeq ({\rm sgn} \mu) \ 130 \times 10^{-11}\ \tan \beta\ 
\left(\frac{100\ {\rm GeV}  }{ \tilde m}\right)^2
\label{eq:blr-susy}
\end{equation}

\subsection{Measurements of the Muon and Electron Anomalies}

The electron anomaly has been measured to a precision of 0.24 parts
billion by storing a single electron in a quantum cyclotron
and measuring the quantum cyclotron and spin levels 
in this system~\cite{blr-hanneke08}.
Were an independent measurement of the
fine-structure constant $\alpha$ available at this precision, this
impressively precise measurement could provide a testing ground for the
validity of QED down to the five-loop level, and present an opportunity to
search for effects of New Physics. At present the best independent
measurements of $\alpha$ have a precision of $\sim 5$~ppb~\cite{blr-alpha}.
  In the absence of such an independent
measurement, the electron $(g-2)$ value has been used, along with the QED
theory (assumed to be valid), to give the most precise 
value of $\alpha$~\cite{blr-hanneke08}. 

The muon anomaly, while only measured to an accuracy of 0.54 parts per
million (ppm)~\cite{blr-bennett04}, nevertheless has an increased 
sensitivity to heavier physics 
that scales as $(m_\mu/m_e)^2 \simeq 43,000$. This means that at a
measurable level the Standard-Model contributions to the muon anomaly come
from QED; from virtual hadrons in vacuum polarization or hadronic light-by-light
scattering loops; and from loops involving the electroweak gauge bosons.

In principle the technique is similar to the measurement of the electron
anomaly,  where muons are stored in a ``trap''
consisting of a dipole magnetic field plus an electrostatic quadrupole
field. In the muon experiment, an ensemble of muons is injected into
a precision storage ring.  The observable is the spin precession
frequency relative to the momentum, which is the difference between the spin
precession frequency and the cyclotron frequency:
\begin{eqnarray}
\vec \omega_a &=& \vec \omega_S - \vec \omega_C \nonumber \\
&=& - \ \frac{Qe }{ m} 
\left[ a_{\mu} \vec B -
\left( a_{\mu}- \frac{1 }{ \gamma^2 - 1} \right)
\frac{ {\vec \beta \times \vec E }}{ c }
\right]\,.
\label{eq:blr-omega}
\end{eqnarray}
The second term in brackets represents the effect of the motional magnetic
field on the spin motion. The experiment is operated at the ``magic'' value
of $\gamma_{magic} = 29.3$ where this motional term vanishes, which permits
the use of an electric quadrupole field to provide the vertical focusing. 

The measured electron and muon anomalies are
\begin{eqnarray}
a_e &=& [115\, 965\, 218\, 073 (28)] \times 10^{-14} \,0.24\,{\rm ppb}\\
a_\mu &=& [116\,592\,089 (63)] \times 10^{-11}\, 0.54\,{\rm ppm}
\label{eq:blr-anomalies}
\end{eqnarray}
Interestingly enough, the muon anomaly seems to be slightly larger than the
Standard-Model value of~\cite{blr-davier09} 
\begin{equation}
a_\mu^{\rm SM}[e^+e^-] = 116\, 591\, 834 (49) ]\times 10^{-11}
\end{equation}
which uses $e^+e^-$ annihilation into hadrons to determine the hadronic
contribution, and the value of Prades et al.,~\cite{blr-PdRV} for the hadronic
light-by-light contribution. There is  a difference of $\sim3.2~\sigma$
between the two.  If hadronic $\tau$ decays are used
to determine the lowest-order hadronic contribution (a determination that
relies on significant isospin corrections) the difference drops to
$\sim 2 \sigma$~\cite{blr-davier-tau09}. 

Such a deviation could
 fit well with the expectations of supersymmetry in the few-hundred GeV mass
 region,  as shown in
Eq.~\ref{eq:blr-susy}.
Were SUSY particles to be discovered 
at LHC, the muon anomaly would play an important role in helping to
discriminate between the different possible scenarios, and providing
a measure of $\tan \beta$. For a thorough 
review of SUSY and $(g-2)$ see the 
articles by St\"ockinger~\cite{blr-stockinger}.

\begin{figure}[h]
\centering
\includegraphics[width=0.4\textwidth]{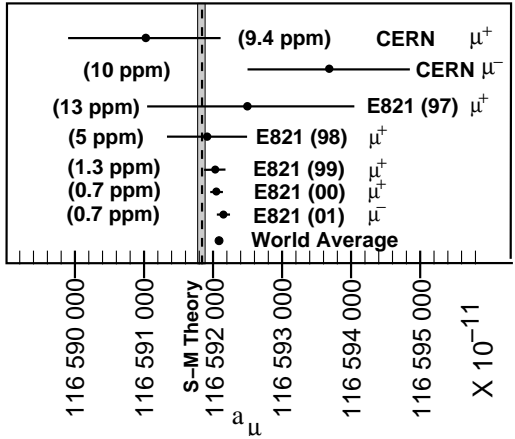}
\caption{Measurements of the muon anomalous magnetic moment. The theory value
shown is taken from Ref.~\cite{blr-davier09} as described in the text.} 
\label{fg:blr-amu}
\end{figure}

The precision of the E821 $(g-2)$ measurement was limited by the statistical
error of 0.46~ppm, compared to the systematic error of 0.28~ppm.  A new
experiment has been proposed for Fermilab, P989~\cite{blr-p989} with the goal of
equal statistical and systematic errors, and a total error of 0.14 ppm, a
factor of four improvement over E821.  

Significant work on different aspects of the
hadronic contribution are in progress, both on the experimental side to
measure the hadronic electroproduction cross sections better, and on theoretical
efforts to improve on the hadronic light-by-light
contribution~\cite{blr-phipsi09}. 

\begin{figure}[h]
\centering
\includegraphics[width=0.4\textwidth]{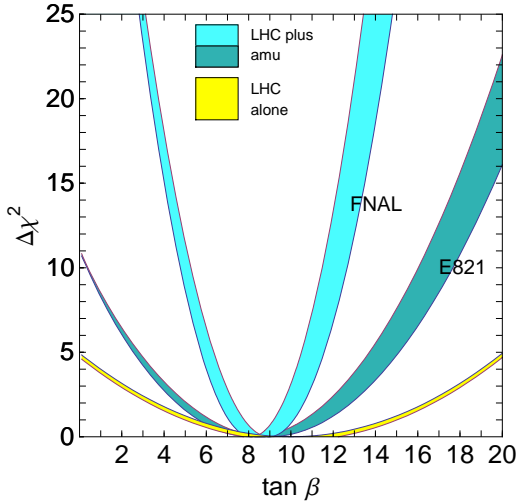}
\caption{Implications for a $\tan \beta$ determination assuming the LHC were
  to discover the SPS1a SUSY scenario, and the $\Delta a_\mu 
= (255 \pm 80) \times 10^{-11}$ using the hadronic contribution from 
$e^+e^-$ collisions given in
 Ref.~\cite{blr-davier09}. (Figure courtesy of Domink St\"ockinger)} 
\label{fg:blr-blueband}
\end{figure}

The supersymmetry community has chosen a number of possible scenarios 
that might be discovered at LHC, the Snowmass points and slopes~\cite{blr-SPS},
which serve as benchmarks for determining the sensitivity to the SUSY
parameters.
Since $a_\mu$ has significant sensitivity to $\tan \beta$ 
(see Eq.~\ref{eq:blr-susy}), it is possible to compare the sensitivity to
$\tan \beta$ from LHC vs. from $\Delta a_\mu$. Such a comparison is shown in
Fig.~\ref{fg:blr-blueband}, which
 assumes that the SPS1a point is realized, 
(typical mSUGRA point with an intermediate value of $\tan \beta$).  There is
some tension between the new value of $\Delta a_\mu$ and this model, which
predicts $\Delta a_\mu = 293 \times 10^{-11} $, so the $\chi^2$ minimum from 
LHC is at 10, the input from SPS1a, and the present $\Delta a_\mu$ value 
implies a slightly lower value.  The lighter blue band shows the improvement
that could be gained in the new Fermilab experiment.

\section{Electric Dipole Moments}

Unlike the magnetic dipole moments, the Standard-Model values of electric
dipole moments are
 orders of magnitude less than present experimental limits, both of which
are shown in Table~\ref{tb:blr-edm}. The the
experimental observation of an EDM would unambiguously 
signify the presence of new
physics.

\begin{table}[h!]
\caption{Measured limits on electric dipole moments, and their Standard
Model values}
\label{tb:blr-edm}
\begin{tabular}{|c|c|c|} \hline
   { Particle}  &{ Present EDM} & { Standard Model}  \\
                 & Limit ($e\cdot$~cm)           & Value ($e\cdot$~cm) \\
\hline
$n$ & {$2.9 \times 10^{-26}$ } (90\%CL)~\cite{blr-nEDM}  & {$\simeq 10^{-32}$ }  \\
\hline
$p$  & $ 7.9 \times 10^{-25}$~\cite{blr-hgEDM} &  {$\simeq 10^{-32}$ }  \\
\hline
 $e^-$  & {$\sim 1.6 \times 10^{-27 }$} (90\%CL)~\cite{blr-eEDM} & {$10^{-38}$ } \\
\hline
{$\mu$} &{$1.8 < 10^{-19}$ } (95\%CL)~\cite{blr-bennett09} & {$10^{-35}$ }\\
\hline
$^{199}{\mathrm Hg}$ & $ 3.1 \times  10^{-29}$  (95\%CL)~\cite{blr-hgEDM} & \\
\hline
\end{tabular}
\end{table}

For hadronic systems, the ``theta'' term in the QCD Lagrangian
\begin{equation}
{\mathcal L }^{eff}_{QCD} = {\mathcal L}_{QCD} + \theta \frac{g^2_{ QCD}}{32 \pi^2}F^{a \mu\nu}
\tilde F_{a \mu \nu }  \ \ a = 1,2,\dots,8
\label{eq:blr-theta}
\end{equation}
violates both parity and time-reversal symmetries,
where the physical quantity is the sum of $\theta$ and the overall phase in
the quark matrix, 
 $\bar \theta = \theta + \arg(\det M)$.
The non-observation of a neutron EDM restricts the value of $\theta$:
\begin{equation}
\left|d_{n}\right|\simeq3.6\times10^{-16}\bar{\theta}\,
e\cdot\mathrm{cm}\ \Rightarrow \ \bar{\theta} \lsim10^{-10},
\label{eq:blr-thetabar}
\end{equation}
which for a quantity that could be order one is anomalously small, and
is often referred to as the strong {\sl CP} problem.  While supersymmetry, 
or other models of New Physics can 
easily contain new sources of {\sl CP} violation, the absence of any observation
 of an EDM, with a significant
 fraction of the ``natural'' part of the SUSY {\sl CP}-violating 
parameter space already eliminated, is sometimes called the SUSY {\sl CP}
 problem.

The isovector and isoscalar combinations of the magnetic dipole
moments are:
\begin{eqnarray}
F_{2N}^{(I=1)} &=& \frac{F_{2p} - F_{2n}}{2} \simeq 1.85 \\ 
F_{2N}^{(I=0)}  &=& \frac{F_{2p} + F_{2n}}{2} \simeq -0.06 \, ;
\end{eqnarray}
we conclude that the isovector dominates the anomalous MDM.  Both isoscalar
and isovector EDMs are predicted by various models~\cite{blr-posp-ritz}, 
so measuring both the
proton and neutron EDMs would help disentangle these two possibilities.

In the traditional EDM experiment, the system is placed in a region of 
parallel (anti-parallel) electric and magnetic fields. The Larmor frequency is
measured, and then the electric field direction is flipped.  An EDM would
cause the Larmor frequency to be higher/lower depending on the direction of
the electric field.  The EDM is determined by the frequency difference
between these two configurations:
\begin{equation}
\Delta \nu =\nu_{\uparrow \uparrow}- \nu_{\uparrow \downarrow} = \frac {4 d E}{h}.
\end{equation} 

A new result from the Seattle group places the limit
on the EDM of the mercury atom~\cite{blr-hgEDM}:
\begin{equation}
d(^{199} {\mathrm Hg}) = (0.49 \pm 1.29_{stat} \pm 0.76_{syst}) 
\times 10^{-29}\, e\cdot{\rm cm}
\end{equation}
giving the limit above in Table~\ref{tb:blr-edm}.

Searches are underway worldwide to find an EDM of the 
electron~\cite{blr-commins} (Imperial College, Colorado, Harvard, 
Yale, Amherst, Penn State,
Texas, Osaka and Indiana),
neutron~\cite{blr-lamoreaux} (ILL, PSI, Oak Ridge), the 
atoms~\cite{blr-fortson} $^{199}$Hg (Seattle) or 
$^{129}$Xe (Princeton), $^{225}$Ra (Argonne, Groningen),

The limit on the muon EDM comes from E821 at Brookhaven~\cite{blr-bennett09}. If
an EDM exists, it is necessary to modify the spin precession formula of
Eq.~\ref{eq:blr-omega} with an extra term, $\omega_\eta$
\begin{equation}
\vec \omega_\eta =
 \eta \frac {Qe}{2m}
 \left[ \frac {\vec{E}} {c}  +  \vec{\beta} \times \vec{B} \right] 
\label{eq:blr-omegaeta}
\end{equation}
and the total spin precession frequency is $\vec \omega = \vec \omega_a +
\vec \omega_\eta$.  The motional electric field is proportional to 
$\vec \beta \times \vec B$, so
 the EDM results in an out-of-plane component of the
spin, where the (very small) tipping angle relative to $\vec \omega_a$ is
$\delta =\tan^{-1} {\omega_{\eta}}/ {\omega_a}
= \tan^{-1} ({ \eta \beta }/{ 2a}).$ 
For spin $1/2$, $\eta$
is related to the EDM, $d$, by the relationship
\begin{equation}
d =  \left ( \frac {e\hbar} {4mc} \right) \eta
\end{equation}

  In the $(g-2)$ experiments, $\omega_\eta \ll \omega_a$ and the resulting
motion is an up-down oscillation with frequency $\omega_a$,
{\em out of phase} with the $(g-2)$
oscillation.  Such an experiment is largely limited by systematic
errors~\cite{blr-bennett09}, since the out-of-plane motion is masked by the 
large-amplitude spin precession from the magnetic moment.
Nevertheless, the new Fermilab effort hopes to achieve one to two orders of
magnitude improvement in the muon EDM as a by-product of the improved $(g-2)$
measurement.  Significant progress beyond that goal 
would need to reduce the large
background caused by the $\omega_a$ precession.  

To achieve this reduction,
the ``frozen spin'' technique has been 
proposed~\cite{blr-farley-fs,blr-roberts-edm}.
 Recall the point of choosing the magic $\gamma$ in
Eq.~\ref{eq:blr-omega} was to eliminate the effect of the focusing electric
field on the spin precession.  If however, a storage ring were to be operated
at a different momentum, then a {\em radial} electric field could be 
used to counter the
 the spin precession from the magnetic moment (see Eq.~\ref{eq:blr-omega},
 {\it viz.}
 it could be chosen such
 that $\omega_a = 0$. 
 The $E$-field required to freeze the muon spin is 
\begin{equation}
E=\frac {a B c \beta \gamma^2} {1 - a \beta^2 \gamma^2} \simeq aBc\beta
\gamma^2  .
  \label{eq:mrs-efield}
\end{equation}
Possible parameters of such an experiment are $E = 2$~MV/m,
 $p_\mu = 500$~MeV/$c$, $\gamma = 5$, 
$R_0= 7$~m~\cite{blr-farley-fs,blr-roberts-edm},
although a much smaller ring has been suggested for the Paul Scherrer
Institut~\cite{blr-kirch}.  The frozen spin technique, along with a very
high-flux facility could permit a sensitivity of $10^{-24}\, e \cdot$cm
 or better for the
muon EDM, providing a unique opportunity to measure the EDM of
 a second generation particle. 

The error on such a measurement is given by~\cite{blr-roberts-edm}
\begin{equation}
\sigma_\eta = \frac{\sqrt 2}{\gamma \tau (e/m)\beta B A \sqrt N},
\label{eq:blr-sigma-dmu}
\end{equation}
which implies that one needs $NA^2 \simeq 10^{16}$ for 
$\sigma_{d_\mu} \simeq 10^{-23}\,e\cdot$~cm. The polarization
enters directly into the asymmetry $A$, thus
the  muon beam for the EDM experiment must  have high polarization.

\begin{figure}[h]
\vskip9mm
\centering
\includegraphics[width=80mm]{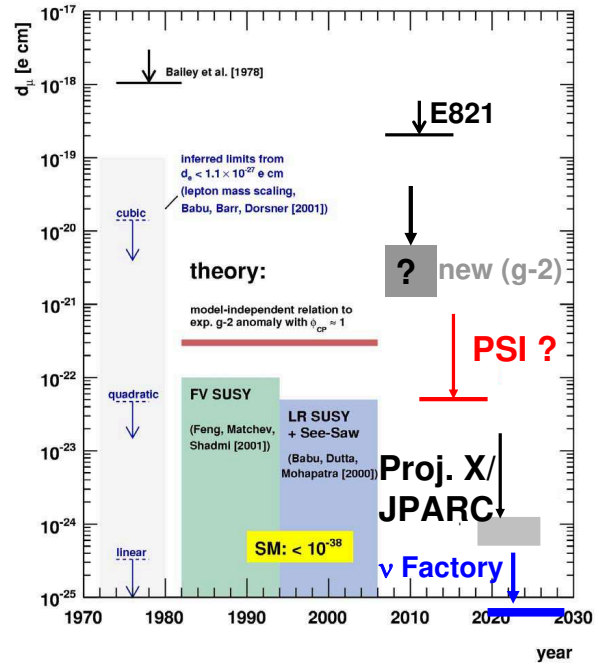}
\caption{Present and projected limits on the muon EDM.
To progress beyond the projection for the new $(g-2)$ experiment
at Fermilab, dedicated storage rings will be required, as 
described in the text. (Modified from a figure
 courtesy of Thomas Schietinger)} 
\label{fg:blr-mu-edm-limits}
\end{figure}

In closing this section, one additional point needs to be made.
Should convincing evidence for any EDM be found, it will be
imperative that as many other EDMs as possible
be measured to help sort out the
source of this new {\sl CP} violation.

\section{Transition Moments}

One of the most important discoveries in the past decade was the 
definitive evidence that neutrinos mix.  In the Standard Model, this implies
that charged leptons will also mix, and when one calculates the 
transition rate for $\mu^+ \rightarrow e^+ \gamma $
one finds:
\begin{equation}
Br(\mu \rightarrow e \gamma)=
\frac{3 \alpha}{32 \pi} \left| \sum_\ell V^*_{\mu \ell} V_{e \ell} \frac{m_{\nu_\ell}^2}{M_W^2}\right| ^2 
\leq 10^{-54}
\end{equation}
which is immeasurable under the most optimistic scenario. 
 Thus the observation of any process that violates lepton flavor would
herald the discovery of new physics.

Just as the diagonal matrix elements of the electromagnetic current
were connected with the electric and magnetic dipole moments, we have
the off-diagonal elements of the current\cite{blr-CzarLM} that give
transition moments:
\begin{equation}
\left\langle
f_{j}\left(p'\right)\left|J_{\mu}^{\mathrm{em}}\right|
f_{i}\left(p\right)\right\rangle
 =  \bar{u}_{j}\left(p'\right)\Gamma_{\mu}^{ij}u_{i}\left(p\right),
\end{equation}
where $\Gamma_\mu^{ij}$ is given by
\begin{eqnarray}
\Gamma_{\mu}^{ij} & = &
\left(q^{2}g_{\mu\nu}-q_{\mu}q_{\nu}\right)\gamma^{\nu}
\left[F_{E0}^{ij}\left(q^{2}\right)
+\gamma_{5}F_{M0}^{ij}\left(q^{2}\right)\right]\nonumber \\
 &  &
 +i\sigma_{\mu\nu}q^{\nu}\left[F_{M1}^{ij}\left(q^{2}\right)
+\gamma_{5}F_{E1}^{ij}\left(q^{2}\right)\right].
\label{eq:2.39}\end{eqnarray}
The first term gives rise to chiral-conserving flavor-changing amplitudes
at $q^2 \neq 0$, e.g. $K^+ \rightarrow \pi^+ e^+e^-$,
$\mu^+ \rightarrow e^+ e^+ e^-$, and the second term gives rise to
chiral-changing, flavor-changing amplitudes, e.g. 
$b \rightarrow s \gamma$, $\mu \rightarrow e \gamma$ and $\tau \rightarrow
e \gamma$.

Here I confine myself to the muon sector, where possible reactions include:
\begin{eqnarray}
\mu^+ &\rightarrow& e^+ \gamma \label{eq:blr-MEG}\\
\mu^+  &\rightarrow& e^+ e^+ e^- \label{eq:blr-M3E} \\
\mu^- {\mathcal N}  &\rightarrow& e^- \label{eq:blr-MEC} {\mathcal N}\\
\mu^+e^-  &\rightarrow& \mu^- e^+ \label{eq:blr-MMBAR}
\end{eqnarray}
 
\begin{figure}[h!]
\centering
\includegraphics[width=80mm]{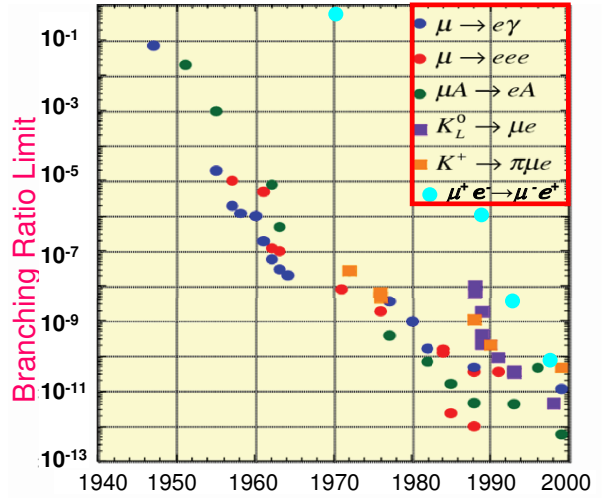}
\caption{The history of searches for muon flavor violation.
} 
\label{fg:blr-limits}
\end{figure}

There is a long experimental history of searches for these reactions, going
back to the search for $\mu \rightarrow e \gamma$ in 1947 by Hinks and
Pontecorvo~\cite{blr-hinks47}, who showed
 the branching ratio was less than 10\%.
In the intervening years, the limits for all these processes have been
lowered to $10^{-10} - 10^{-12}$, as shown in Fig.~\ref{fg:blr-limits}.
  Ambitious experiments planned or in
preparation have goals of $10^{-18}$ or below.

\begin{figure}[h]
\centering
\includegraphics[width=80mm]{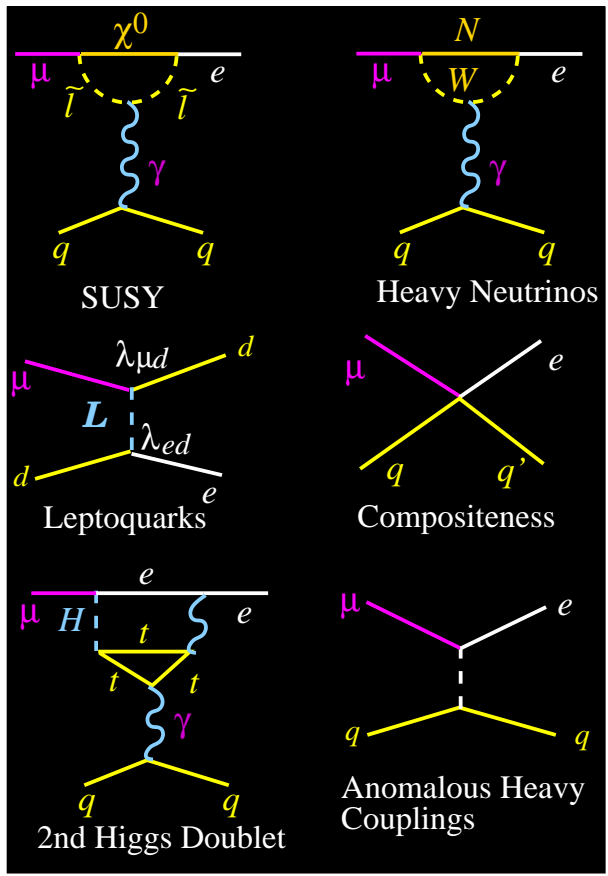}
\caption{Diagrams that might contribute to muon-electron conversion.
(Following W. Marciano)} 
\label{fg:blr-MEC}
\end{figure}

 A wide range of NP could produce such transitions. 
One channel which permits the highest experimental sensitivity is
the muon to electron conversion reaction, Eq.~\ref{eq:blr-MEC}.
If negative muons are stopped in matter, they come to rest and get captured
into atomic orbits in the stopping material. They
 then cascade down to the atomic $1s$ state. Ordinarily the $\mu^-$
either decay in orbit, or are captured weakly on the atomic nucleus,
which is analogous to K capture of atomic electrons.
In the coherent conversion to an electron with no neutrinos, the 
 signal is a mono-energetic electron with an energy equal to the muon
mass less the atomic binding energy of the muon in the ground state of
the muonic atom.  While this process is forbidden in the Standard Model,
it is possible in a large number of Standard-Model extensions,
some of which are shown diagrammatically in Fig.~\ref{fg:blr-MEC}

\begin{figure}[h]
\centering
\includegraphics[width=80mm]{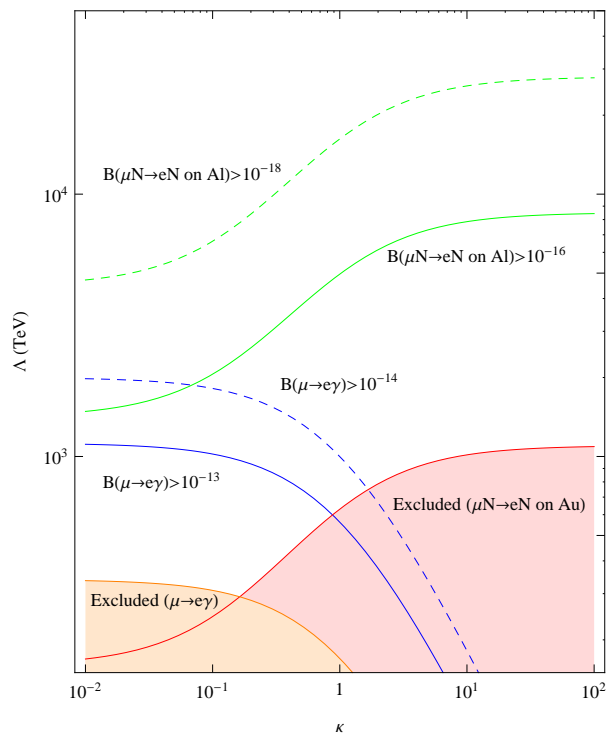}
\caption{The mass reach of the CLFV reactions as a function of the strength
  of
the non-dipole terms. The green curves show the sensitivity of the 
Mu2e experiment a Fermilab, the solid being for the phase using the present
booster accelerator, and the dashed being for the ``Project X'' era.
(Figure courtesy of Andrew Norman)} 
\label{fg:blr-mass-limits}
\end{figure}

The muon-electron conversion (MEC)
 is especially interesting because of the broad
range of physics which it addresses. The interaction Lagrangian ${\mathcal L}_{\rm int}$
is given by~\cite{blr-kitano}:
\begin{eqnarray}
&&  - {4G_F \over \sqrt{2} } \left( 
m_\mu A_R \bar \mu \sigma^{\mu \nu} P_L e F_{\mu \nu} + m_\mu A_L \bar \mu \sigma^{\mu \nu}P_R e F_{\mu \nu} + 
 \right) \nonumber  \\
 &&- {G_F \over \sqrt {2} }  \sum_{q=u,d,s} \left[ 
\left( g_{LS(q)} \bar e P_R \mu + g_{RS(q)} \bar e P_L \mu \right) \bar q q \ 
 \right] \nonumber \\
&&- {G_F \over \sqrt {2} }\left[
  \left( g_{LV(q)} \bar e\gamma^\mu  P_R \mu + g_{RV(q)}  \bar e \gamma^\mu P_R \mu \right)\bar q \gamma_\mu q    
\right] \nonumber  \\
&& + {\rm h.c.\ for \ each \ term} 
\end{eqnarray}
where the three terms in the Lagrangian represent
dipole, scalar and vector interactions respectively.  If the dipole
dominates, then muon-electron conversion is suppressed by 
$2\, - \,4 \times 10^{-3}$ relative to 
$\mu \rightarrow e \gamma$, however, for the other operators, MEC is
much more sensitive.  This is illustrated in Fig.~\ref{fg:blr-mass-limits},
where the sensitivity to different mass scales $\Lambda$ is shown as a
function of the amount of the non-dipole contribution
$\kappa$~\cite{blr-marciano}.  While 
$\mu \rightarrow e \gamma$ is much more sensitive if the dipole contribution
dominates, for non-dipole interactions, the conversion
experiment has an enormous mass reach, well beyond what could be imagined at
colliders.  The muon-electron conversion rate depends both on the operator,
and on the nucleus~\cite{blr-kitano}.  If it 
becomes possible to measure MEC for a range of nuclei, the $Z$ dependence will
help disentangle which operators are responsible.  Furthermore, the observation
of several CLFV processes will help further, perhaps along with the electric
and magnetic dipole moment information.

At present there is one running experiment in the muon sector, the MEG
experiment at the Paul Scherrer Institut, which aims for a sensitivity of
$R_{\mu e \gamma} \sim 1 \times 10^{-13}$ for the 
process $\mu^+ \rightarrow e^+ \gamma$. 
Of course, experimentally, this channel is quite challenging.  The two-body
final state uniquely determines the kinematics, so at rest the photon and
electron are back-to-back, sharing the muon mass energy. However,
photons in the 50~MeV energy
region are difficult to detect with good position and directional
 information on
the photon.  A preliminary result from MEG reported
a $3\times 10^{-11}$ 90\% confidence level limit~\cite{blr-MEG}, 
which is not yet
competitive with the present limit of $1.2 \times 10^{-11}$~\cite{blr-mega}.

The muon-electron conversion experiment is rather special, since the signal
of a single mono-energetic electron is unique, and in principle resolved from
background. The two proposals to study this process, Mu2e at Fermilab
which advertises a first phase 90\% CL limit $R_{\mu e}< 6 \times 10^{-17}$, and
COMET at J-PARC which proposes
 to reach a sensitivity of $< 10^{-16}$ for its first
phase. The sensitivity of the
 second phase of these experiments is projected to be $\sim 10^{-18}$
The Mu2e experiment at Fermilab has stage-one
approval, and significant engineering work is ongoing. 
The COMET experiment at J-PARC is still under review by the J-PARC Laboratory.
See the reviews~\cite{blr-kuno,blr-kuno-okada,blr-okada}, 
and references therein,
for a detailed discussion of 
charged lepton flavor violation
experiments and their physics reach, as well as for additional details on COMET.

\section{Conclusions}

I have described a set of experiments from the precision/intensity frontier
which have the potential impact equal to the discoveries we hope for, and
expect to find at the LHC.  The discovery of a permanent electric dipole
moment would herald, at long last, a new source of {\sl CP}
 violation that might
 explain the matter-antimatter asymmetry of the universe, and
partially explain why we are here.  The discovery of charged lepton flavor
violation would also herald New Physics at work in the lepton sector.  A
confirmation of the muon $(g-2)$ discrepancy would also signify new physics
at the loop level.  All of these experiments will help guide our
interpretation of the new phenomena which we hope to discover at LHC.
Perhaps the most important message from this talk is that many different
additional  
experimental results will be necessary to
help guide our interpretation of the discoveries 
made at the LHC.  It is crucial for the future health
 of the field that a diverse program, exploring both the precision and energy
 frontiers, be strongly supported.

\begin{acknowledgments}
I wish to thank Gerco Onderwater for his careful reading of this manuscript,
and for many useful suggestions.
I wish to acknowledge Andzrej Czarnecki, Michel Davier, David Hertzog,
Yoshi Kuno, Bill Marciano, Jim Miller, 
Eduardo de Rafael, Yannis Semertzidis and Graziano Venanzoni,
for helpful conversations.  
The review article by Czarnecki and
Marciano in Lepton Dipole Moments~\cite{blr-LM} was extremely 
useful in preparing
this talk.  Thanks to Thomas Schietinger for providing the core of 
Fig.~\ref{fg:blr-mu-edm-limits}, and to
Andrew Norman for providing Fig.~\ref{fg:blr-mass-limits}.

\end{acknowledgments}

\bigskip 

\end{document}